\documentclass{emulateapj}

\newcommand{\jmass}{ $\rm J_{2M}\;$}

\newcommand{\jnf}{ $\rm J_{NF \;} $}

\newcommand{\lya}{ Ly$\alpha \;$}
\def\ergcm2s{\ifmmode {\rm\,erg\,cm^{-2}\,s^{-1}}\else
                ${\rm\,ergs\,cm^{-2}\,s^{-1}}$\fi}

\newcommand{\ha}{ $\rm H{\alpha}\;$}
 
\slugcomment{Published in ApJ}

\shorttitle{The Luminosity Function of Lyman alpha Emitters at Redshift $z=7.7$}
\shortauthors{Tilvi et al.}

\begin{document}

\title{The Luminosity Function of Lyman alpha Emitters at Redshift $z=7.7$}

\author{Vithal Tilvi\altaffilmark{1}, 
James E. Rhoads\altaffilmark{1},
Pascale Hibon\altaffilmark{1},
Sangeeta Malhotra\altaffilmark{1},  
Junxian Wang\altaffilmark{2},
Sylvain Veilleux \altaffilmark{3}, 
Rob Swaters \altaffilmark{3},
Ron Probst  \altaffilmark{4},
 Hannah Krug \altaffilmark{3},
 Steven L. Finkelstein  \altaffilmark{5}
 and Mark Dickinson \altaffilmark{4}
 }

\altaffiltext{1}{School of Earth and Space Exploration, Arizona State University,  Tempe, AZ 85287, USA ; tilvi@asu.edu}
\altaffiltext{2}{Center for Astrophysics, University of Science and Technology of China,  Anhui 230026, China}
\altaffiltext{3}{Department of Astronomy, University of Maryland, College Park, MD 20742, USA.}
\altaffiltext{4}{NOAO, Tucson, AZ 85719, USA. NOAO is operated by the Association of Universities for Research in 
Astronomy (AURA), Inc., under cooperative agreement with the National Science Foundation.}
\altaffiltext{5}{Texas A\&M University, College Station, TX. }

\begin{abstract}
  Lyman alpha (\lya) emission lines should be attenuated in a neutral
  intergalactic medium (IGM).  Therefore the visibility of \lya\
  emitters at high redshifts can serve as a valuable probe of
  reionization at about the 50$\%$ level.   We present an imaging search for $z=7.7$ \lya\
  emitting galaxies using an ultra-narrowband filter (filter width=$\rm 9\AA$)
   on the NEWFIRM imager at the Kitt Peak
  National Observatory.  We found four candidate \lya\ emitters in a
  survey volume of $1.4\times 10^{4} \rm Mpc^{3}$, with a line flux
  brighter than $6\times 10^{-18} \ergcm2s$ (5$\sigma$ in 2$^{\prime \prime}$
    aperture). We also performed a detailed Monte-Carlo simulation
  incorporating the instrumental effects to estimate the expected
  number of \lya\ emitters in our survey, and found that we should
  expect to detect one \lya\ emitter, assuming a non-evolving \lya\ luminosity function (LF)
  between $z$=6.5 and $z$=7.7. Even if one of the present candidates
  is spectroscopically confirmed as a $z\approx 8$ \lya\ emitter, it
  would indicate that there is no significant evolution of the \lya\
  LF from $z=3.1$ to $z\approx 8$. While firm conclusions would need
  both spectroscopic confirmations and larger surveys to boost the
  number counts of galaxies, we successfully demonstrate the feasibility
  of sensitive near-infrared ($1.06 \mu$m) narrow-band searches using custom
  filters designed to avoid the OH emission lines that make up most of the
  sky background.
  \end{abstract}


\section{Introduction}

Lyman alpha (\lya) emitting galaxies offer a powerful probe of both
galaxy evolution and the reionization history of the universe.  \lya\
emission can be used as a prominent signpost for young galaxies whose
continuum emission may be below usual detection thresholds. It is also a
tool to study their star formation activity, and a handle for
spectroscopic followup.  

The intergalactic medium (IGM) will obscure \lya\ emission from view
if the neutral fraction exceeds $\sim 50\%$ \citep{fur06,mcq07}.  
%% Good to cite McQuinn here, and Furlanetto 
Recently, \lya\ emitters
have been used to show that the IGM is $\lesssim 50\%$ neutral at
$z=6.5$ \citep{rho01,mr04, ste05, kas06,mr06}.
%%{\it cite Rhoads & Malhotra 2001, Stern et al. 2005, Kashikawa et   al. 2006}.  
This complements the Gunn-Peterson $lower$ bound of
$x_{HI} \gtrsim 1 \%$ at $z\approx 6.3$. Completely independently,
polarization of the cosmic microwave bakground suggests a central
reionization redshift $z_{re}=10.5 \pm 1.2$ \citep{kom10}.

In addition to their utility as probes of reionization,  \lya\ emitters 
are valuable in understanding galaxy formation and evolution at the
highest redshifts. 
This is especially true for low mass galaxies, as \lya\ emitters are
observed to have stellar masses $\rm M_{\star} \lesssim 10^{9} \;
M_{\odot}$ \citep{gaw06, pir07, fin07,pen09}, appreciably below
the stellar masses of Lyman break selected galaxies (LBG) \citep{ste96} 
at similar redshifts \citep[e.g.][]{pap01, sha01, sta09}.

Narrow-band imaging is a well established technique for finding 
high redshift galaxies \citep[e.g.][]{rho00a, rho04, rho03, mr02,
  mr04, cow98, hu99, hu02, hu04, kud00, fyn01, pen00, ouc01, ouc03,
  ouc08, sti01, shi06, kod03, aji04, tan05, ven04, kas06, iye06,
  nil07, fin09}.  The method works because \lya\ emission redshifted
into a narrow band filter will make the emitting galaxies appear brighter
in images through that filter than in broadbands of similar wavelength.
A supplemental requirement that the selected emission line galaxies be
faint or undetected in filters blueward of the narrowband filter
effectively weeds out lower redshift
emission line objects \citep[e.g.][]{mr02}.  This has proven to be very efficient
for selecting star-forming galaxies up to $z\lesssim 7$, and remains effective
even when those galaxies are too faint in their continuum emission to 
be detected in typical broadband surveys.

While large samples of \lya\ emitters have been detected at $z<$6,
both survey volumes and sample sizes are much smaller 
at $z>$6.  Since the \lya\ photons
are resonantly scattered in neutral IGM, a decline in the observed
luminosity function (LF) of \lya\ emitters would suggest a change in the
IGM phase, assuming the number density of newly formed galaxies remains
constant at each epoch.  \citet{mr04} found no significant evolution
of \lya\ LF between $z$=5.7 and $z$=6.6, while
\citet{kas06} suggested an evolution of bright end of the \lya\ LF in
this redshift range.  At even higher redshifts, $z=6.5$ to $z$=7, some
authors \citep{iye06, ota08} suggest an evolution of the \lya\ LF however
based on a single detection.
 
Recently, \citet{hib09} found seven \lya\ candidates at $z$=7.7 using
the 
 Wide-Field InfraRed Camera on the 
 Canada- France-Hawai`i Telescope.  
If these seven candidates are real and high redshift galaxies, the derived \lya\ LF 
suggest no strong evolution from $z$=6.5 to $z$=7.7. 
\citet{sta07} found six candidate \lya\ emitters at $z\approx 8 -10$ in a spectroscopic
survey of gravitationally lensed \lya\ emitters.
 Other searches
\citep[e.g.][]{par94, wil05,cub07, wil08, sob09} at  redshift $z\gtrsim$8  either had insufficient volume
or sensitivity, and hence did not find any \lya\ emitters.

In this paper we present a search for \lya\ emitting galaxies at
$z=$7.7, selected using custom-made narrowband filters that avoid
night sky emission lines and therefore are able to obtain low
sky backgrounds.  This paper is organized as follows. In
section~\ref{sec:data}, we describe in detail the data and reduction.
In section~\ref{sec:selection} we describe our selection of \lya\
galaxy candidates.  In section~\ref{sec:contam} we discuss possible
sources of contamination in the sample, and our methods for minimizing
such contamination.  In section 5 we estimate the number of \lya\ 
galaxy candidates expected in our survey using a full Monte Carlo
simulation. In section 6 we discuss the \lya\ luminosity function,
and in section 7 we  compare the \lya\ equivalent widths with previous work.
We summarize our conclusions in section 8. 
Throughout this work we assumed a flat $\Lambda$CDM cosmology with parameters 
$\Omega_{m}$=0.3, $\Omega_{\Lambda}$=0.7, 
$h$=0.71  where $\Omega_{m}$, $\Omega_{\Lambda}$, and 
 $h$  correspond, respectively to the matter density,
dark energy density  in units of the critical density, and 
the Hubble parameter in units of 100 km s$^{-1}$ Mpc$^{-1}$.
All magnitudes are in AB magnitudes unless otherwise stated.

\section{Data Handling} \label{sec:data}

\subsection{Observations and NEWFIRM Filters}
We observed the Large Area Lyman Alpha  survey (LALA)
Cetus field (RA 02:05:20, Dec -04:53:43)  \citep{rho00b} during a six night observing
run with the NOAO\footnote{National Optical Astronomy Observatory}
 Extremely Wide-Field Infrared Mosaic (NEWFIRM) imager \citep{aut03} at the 
Kitt Peak National Observatory's 4m Mayall Telescope during October 1-
6, 2008.  

We used the University of Maryland 1.063 $\mu m$ ultra-narrowband (UNB)
filter, for  a total of 28.7 hours of integration time,
 along with 5.3 hours' integration in the broadband
J-filter.  Both narrow- and J-band data were obtained on each clear
night of observing.  NEWFIRM covers a 28$^{\prime}$ $\times$
28$^{\prime}$ field of view using an array of four detector chips
arranged in a 2$\times$2 mosaic, with adjacent chips separated by a
gap of 35$^{\prime \prime}$.  Each chip is a 2048$\times$2048 pixel
ALADDIN InSb array, with a pixel scale of 0.4$^{\prime \prime}$ per
pixel.  The instantaneous solid angle coverage of the NEWFIRM camera
is about $745 \square^\prime$.

The LALA Cetus field has been previously studied at shorter
wavelengths, most notably by the LALA survey \citep{mr02,wan09}
in narrow
bands with $\lambda_c \approx $656, 660, 664, 668, and 672 nm, and
$\Delta \lambda \approx 80$\AA); the NOAO Deep Wide Field Survey
(NDWFS) \citep{jan99}, with broadband optical B$_w$, R, and I filters;
using MMT/Megacam g$^{\prime}$, r$^{\prime}$, i$^{\prime}$ and z$^{\prime}$ filters
\citep{fin07}; and Chandra, with 180 ksec of
ACIS-I imaging \citep{wan04, wan07}.  In summary, we use narrow-band
UNB \& broadband J data obtained using NEWFIRM, and previously
obtained B$_w$, R, and I -band data (NDWFS) for this study.  
The MMT/Megacam images cover about $55\%$ of the area we observed with NEWFIRM, and
we used these deeper optical g$^{\prime}$, r$^{\prime}$, i$^{\prime}$ and z$^{\prime}$ 
images \citep{fin07} to check our final \lya\ candidates where possible (see section 3 below).

The J filter on NEWFIRM follows the \citet{tok02} filter specifications, 
with $\lambda_c = 1.25\mu m$ and a FWHM of $0.16 \mu m$.
The  UNB filter is an ultra narrow-band filter, similar to the DAzLE narrow-band filters \citep{hor04},  centered at
 1.063  $\mu m$ with a full width at   half maximum (FWHM)  of 8.1 $\rm \AA$.  
We used Fowler 8 sampling (non-destructive readout) in all science frames.  In the UNB filter, we used single 
1200 second exposures between dither positions; in the J band, two coadded 30 second frames.

The NEWFIRM filter wheel places the filters in a collimated beam.  As a consequence,
the effective central wavelength of the narrowband filter varies with position in the
field of view.   Beyond a radius of $12^\prime$, the central wavelength of the 
UNB filter shifts sufficiently to include two weak OH emission lines in the bandpass,
which appear as concentric rings in the narrowband images, and which
limit the survey area where the filter's maximum sensitivity (limited by only the inter-line sky 
background) can be achieved.  
Figure 1 shows the narrowband filter transmission curve along with night sky OH emission lines. 
The UNB filter is designed to avoid OH lines.

\begin{figure}[t!]
\epsscale{1.1}
\plotone{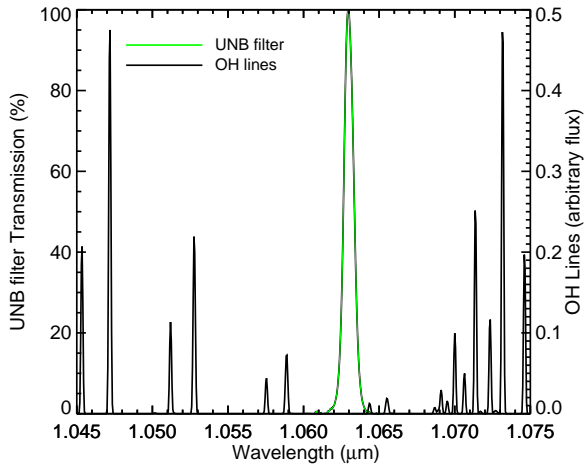}
\caption{ Normalized   narrowband filter transmission curve (green line) and night-sky OH emission lines 
\citep{rou00} (black line) with arbitrary flux. 
Here we have shown the  transmission curve (at the center of the field) of only narrowband filter 
 to demonstrate the use of very narrow region between OH lines, to search for \lya\ 
emitters at $z$=7.7. Two weak OH emission lines with $\lambda = 1.05888$ and $1.05754\mu m$
\citep{rou00} appear in the UNB images  as  concentric rings beyond 12' radius, since the central wavelength of the UNB filter shifts to the blue for positions away from field center.
}
\end{figure}

\subsection{Data  Reduction}
We reduced UNB \& J-band data  using a combination of standard IRAF\footnote{
The Image Reduction and Analysis Facility (IRAF) is distributed by the NOAO, which is operated 
by the Association of Universities for Research in Astronomy, Inc. (AURA) under the cooperative
agreement with the National Science Foundation.} tasks, predominantly
from the {\it mscred} \citep{val98} and  {\it nfextern}\footnote{An external IRAF package for  
NEWFIRM data reduction} \citep{dic09} packages, 
along with custom IDL\footnote{Interactive Data Language} reduction procedures. 

To remove OH rings from UNB data, we created a radial profile for each individual exposure, 
smoothed over a small radius interval $dr$, and subtracted this profile from the exposure.
We then performed sky subtraction by median averaging two OH ring-subtracted frames that were taken 
immediately before and two  frames after the science frame in consideration.  
We then performed cosmic ray rejection on sky-subtracted frames, using the algorithm of \citet{rho00a}.
Prior to  flat-fielding performed using dome-flats, we created a bad pixel mask for each science frame 
by combining the cosmic ray flagged pixels with a static bad pixel mask for the detector.
We then replaced all bad pixels with zero (which is the background level in these sky-subtracted
images) prior to any resampling of the frames. 
We adjusted the  World Coordinate System of individual frames by matching the point sources to 
the 2MASS point 
source catalog using IRAF task $msccmatch$.   % we should state the interpolant here, ideally. Done
We then combined the four chips (i.e four extensions) of each science exposure into a single 
simple image using the $mscimage$ task in IRAF, which interpolates the data onto a common 
pixel grid.
Here we used $sinc17$ for the interpolation.
Using \textit{mscstack} in IRAF, we then stacked all of the narrowband exposures into a single, final narrowband stack.  Pixels flagged as bad are omitted from the weighted averages in this step.
The average FWHM of our final narrowband stack was 1.36$\arcsec$.
In addition to this stacked image, we also generated individual night stacks, which were later used to
identify glitches in \lya\ candidate selection.

For broadband J-filter data reduction, we followed essentially the same procedure,
modified by omission of the OH ring subtraction which is rendered unnecessary by
the absence of noticeable OH rings in the much broader J bandpass.

We now assess the accuracy  of sky subtraction method, the distribution of noise, and the 
uncertainty in  the astrometric calibration of the UNB and J-band stacks.
To evaluate the sky subtraction, and to understand the noise distribution
we constructed sky background, and background 
noise maps using SExtractor  \citep{ber96}. 
The sky subtraction is sufficiently  uniform throughout the image 
except  in the corners  i.e. beyond the OH lines affected regions.
The noise, due to sky brightness,  is also consistent with the expected Poisson noise distribution from sky photons.

To evaluate the uncertainty in the astrometric calibration, we compared
the world coordinates of the sources in the UNB stack (obtained using 
SExtractor) and the corresponding object coordinates from the 2MASS 
catalog. 
We found that the uncertainty in the astrometric calibration is very small, and 
independent of the position in the UNB image.
The  rms of the matched coordinates of UNB and 2MASS is about
0.2 and 0.3 arcseconds corresponding to RA and DEC respectively.

We obtained reduced stacks of deep optical broadband data in B$_w$, R, and I filters, previously observed by the NOAO Deep Wide Field Survey (NDWFS).

At the end,  we have one deep UNB stack, along with five single-night UNB stacks,  
 four broadband stacks in J,   B$_w$, R, and I filters, 
 and four deep stacks in 
  g$^{\prime}$, r$^{\prime}$, i$^{\prime}$ and z$^{\prime}$ \citep{fin09}.
All the stacks were then geometrically matched for ease of comparison.

\subsection{Photometric Calibration}
We performed photometric calibration of UNB \& J-band $\rm(J_{NF})$ data by comparing unsaturated point sources,
extracted using SExtractor, with 2MASS stars.
From 2MASS catalog we selected only those stars that had J-band $\rm(J_{2M})$ magnitudes between 
13.8 \& 16.8 AB mag\footnote{ Since $\rm J_{2M}$ magnitudes are in Vega, we adopted the following conversion between Vega and AB 
magnitudes :  $J_{AB}$=$\rm J_{2M}$ + 0.8 mag},
and 
errors less than 0.1 magnitude.
Since four quadrants of the UNB stack had slightly different zero-points, 
we scaled three quadrants, selected geometrically,  to the 
fourth quadrant, which was closest to the mean zeropoint, by multiplying each quadrant with suitable scaling factors 
so as to make zero-point uniform
throughout the image.
We then obtained zero-points for UNB and \jnf by minimizing the difference between UNB \& \jmass,
and between \jnf \& \jmass respectively.
This left 0.09 rms mag between \jmass \& \jnf magnitudes, and 0.07 rms mag between UNB \& \jmass magnitudes.
The photometric calibration was based on about 30 \&  80  2MASS stars for  narrow-band and J-band respectively.
So the accuracy of the photometric zero points is about $\pm 0.02$
mag in both J and UNB filters.

In addition to the error we have already estimated, there is some uncertainty
	arising due to  different filter widths, and differing central wavelengths of the  2MASS and UNB filter. 
	To estimate
	this uncertainty we constructed observed spectral energy distribution (SED) 
	of stars that were common to both, the 
	 UNB image and   $\rm g^{\prime}, \; r^{\prime}, \; i^{\prime}, \; z^{\prime},  \; J, \; H, \;  \& \;  K$  images. 
From each SED linearly interpolated flux at central wavelengths of the UNB filter and J-filter
were measured.
	From these SEDs we found the median offset between the UNB and J band to be $<$ 0.1 mag. 
	This residual color-term uncertainty in the photometric zero points is smaller than the photometric flux 
	uncertainty in any of our \lya\ candidates.

Before we proceed to calculate the limiting magnitudes, 
we estimate the sky brightness between the OH lines in the 
 UNB image. To estimate this sky value, we construct the UNB stack
 in the same way as described  in Section 2.2 but omitting the OH  
 ring subtraction, and sky subtraction.
 In addition, we subtracted dark current counts from each raw frame.
We estimated the average sky brightness in the UNB image by selecting  
30 random regions avoiding astronomical objects 
 and OH rings.
 This gives us the sky brightness, between the OH lines,   of about 
  21.2  $\rm mag \;arcsec^{-2}$ equivalent to 162 photons $\rm s^{-1} \; m^{-2} \; arcsec^{-2} \; \mu m^{-1}$. 
 This sky brightness is much fainter than the J-band sky brightness which is about   16.1  $\rm mag \;arcsec^{-2}$
equivalent to 17000 photons $\rm s^{-1} \; m^{-2} \; arcsec^{-2} \; \mu m^{-1}$  \citep{mai93} .
 However, more careful analysis are needed to 
   estimate the interline sky brightness in the UNB images.

\subsection{Limiting Magnitudes}
To obtain limiting magnitudes of stacked images, we performed a series of artificial source simulations.
In each, we introduced 400 artificial point sources in an 0.1 magnitude bin of flux 
in the final stacked image.  The positions were chosen randomly, but constrained to avoid
places close to bright stars and already existing sources.
We then ran SExtractor, with the same parameters as were used for the real source detection (see Section 3), 
to calculate the fraction of recovered artificial  sources. 
We ran 20 such simulations in each 0.1 magnitude bin from UNB = 21 to 24 mag.
The 50$\%$ completeness level is UNB$ =  22.5$ mag, which corresponds to an
emission line flux of $6\times 10^{-18} \ergcm2s$.   The very narrow bandpass results in
a relatively bright continuum limit (compared to more conventional 
narrowband filters with $1\%$ to $1.5\%$ bandpass), but the conversion between
narrowband magnitude and line flux is extremely favorable, so that our line flux limits are
competitive with any narrowband search in the literature.
The  50$\%$ completeness for other filters B$_w$, R, I, and \jnf correspond to 26.3, 25.4, 25.0, and
23.5 mag respectively.

\section{\lya\ candidate selection} \label{sec:selection}
We identified sources in the stacked narrowband image using SExtractor.
To measure their fluxes at other wavelengths, we  first combined the broad-band optical 
images B$_w$, R, I into a single 
chi-squared image \citep{sza96} constructed using \textit{Swarp}\footnote
{$Swarp$ is a software 
program designed to resample and combine FITS images.}\citep{ber02}.
A chi-square image is constructed  using the probability distribution of sky pixels in each of the
images to be combined, and extracting the pixels that are dominated by object flux.
We then used SExtractor in dual-image mode in order to measure object fluxes
in both the broad J filter and the combined optical chi-squared image.  In dual
image mode, a {\it detection image} (UNB in this case) is used to identify the pixels associated with each object,
while the fluxes are measured from a distinct {\it photometry image}.

To identify \lya\ candidates in our survey, we used the combined
optical image, UNB image, and J-band image. 
Each \lya\ candidate had to satisfy all the following criteria: 
\begin{itemize}
\setlength{\itemsep}{0cm}%
  \setlength{\parskip}{0cm}%
  \item[(a)]  $5\sigma$ significant detection in the UNB filter,
\item[(b)]  $3\sigma$ significant narrowband excess (compared to the J band image),
\item[(c)]  flux density ratio $f_\nu(UNB) / f_\nu(J) > 2$, 
\item[(d)] non-detection in the combined chi-square optical image (with $<2\sigma$
significance),   
\item[(e)]  consistent with constant flux from night to night (see Section 3.1), and 
\item[(f)]  non-detection in individual optical images.
\end{itemize}
Criteria {\it a-c} ensure real emission line sources. Criterion {\it d} eliminates most low-redshift sources, 
{\it e} eliminates time  variable sources and other glitches, and 
criterion {\it f} eliminates LBGs at $z \ga 4$ which might show up
more clearly in the R or I band than in the $\chi^2$ image
We also used deeper optical  images \citep{fin07} in the overlapping field between MMT/Megacam
and NEWFIRM for criterion {\it f}.

The  criteria follow the successful searches for \lya\ galaxies at lower
redshifts of z=4.5 and 5.7, which have  $\approx 70-80\%$ spectroscopic
success rate \citep{rho01, rho03,daw04,daw07, wan09}.

\begin{figure*}[t!]
\epsscale{0.6}
%{ \hspace{1cm}combined optical }{ \hspace{2cm}UNB   \hspace{2.5cm}   J \hspace{0.1cm}}\\
\plotone{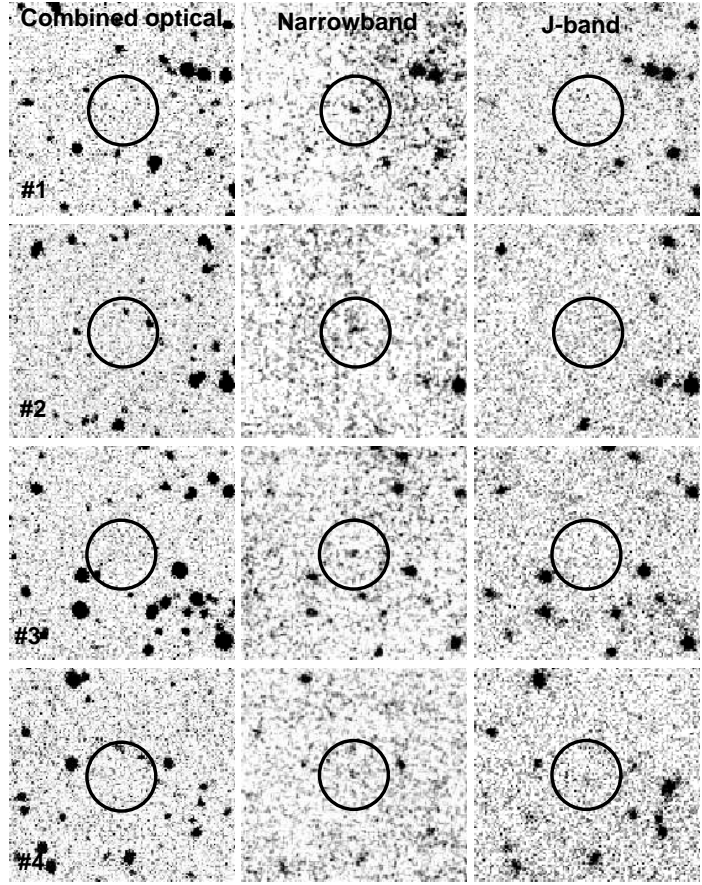}
\caption{ Postage stamps (50$^{\prime \prime}$ wide) of all four \lya\ candidates  in combined (chi-square)
optical image (left panel), UNB (middle), and J-band filter (right panel).
The positions of 
\lya\ candidates are marked with circles 16$^{\prime \prime}$ in diameter.}
\end{figure*}

\subsection{Constant flux test}
In our constant flux test (criterion {\it e} above), we looked at the variation
of flux of each \lya candidate over five nights.
We reject any source having individual night stack fluxes close to zero or showing flux variations  
above a certain chi-square value.
To do this, we generated light-curves of each candidate using individual night stacks of UNB, and then 
determined the $\chi^2$ of the data with respect to the best-fitting constant flux.
Since we had five nights of data, we selected only those candidate that had a chi-square $<$5.
This is in addition to requiring $s/n > 5$, which guards against  peaks in the sky noise 
entering the candidate list.

We also eliminated all the sources that were very close to the chip boundaries.
Combining these criteria with the set of criteria from Section 3, 
we had six \lya\ emitter candidates. 
To increase the reliability of these candidates, we finally selected four candidates  after independent visual inspection 
by four of the authors.
Figure 2 shows postage stamps of all four \lya\ candidates. The candidates are clearly visible in the 
UNB images (middle panel), while undetected in the combined optical (left panel), and J band images (right panel).
We provide the coordinates of our \lya\ candidates in Table 1.

\begin{center}
\begin{table}[ht]
\caption{Coordinates of  our \lya\ candidates.} 
{\small
\hfill{}
\begin{tabular}{l*{3}{c}r}
\tableline
		& RA(J2000)				& DEC (J2000)		\\
LAE 1	& 02:04:45.9				& -04:53:00.8		\\
LAE 2	& 02:04:41.3				& -05:00:11.5		\\
LAE 3	& 02:04:53.2				& -04:46:43.8		\\
LAE 4	& 02:05:58.0				& -05:05:48.4		\\
\hline
\end{tabular}}
\hfill{}
\label{tb:tablename}
\end{table}
\end{center}

\section{Contamination of the sample}\label{sec:contam}
While we have carefully selected \lya\ candidates based on photometric and  geometric  criteria, it
is possible that our \lya\ candidates can be contaminated  by  sources that include  transient objects
such as supernova, cool stars (L \& T dwarfs), foreground emission line sources, and 
electronic noise in the detector.
We now discuss  the possible contribution of sources that can contaminate our \lya\ candidate 
sample.

\subsection{Foreground emission line objects}
Our \lya\ candidate selection can include foreground emission line sources  including
[O\,{\sc ii}]
emitters ($\lambda=3727 \rm \AA$) at $z$=1.85, [O\,{\sc iii}] ($\lambda=5007 \rm \AA$) emitters 
at $z$=1.12, and H$\rm \alpha$($\lambda=6563 \rm \AA$) emitters at $z$=0.62, if they have
strong  emission line flux but faint continuum emission.
We now estimate the number of foreground emitters  that can pass our \lya\ candidate selection criteria.

In our UNB stack, the 50\% completeness limit corresponds to a flux of $6\times10^{-18} \rm erg \; s^{-1}\; cm^{-2}$.
Therefore the minimum luminosities required by the foreground emission line sources to be detected in our
survey are   $1.5\times10^{41} \rm erg \; s^{-1}$,   $4\times10^{40} \rm erg \; s^{-1}$, and 
$1\times10^{40} \rm erg \; s^{-1}$  for  [O\,{\sc ii}],  [O\,{\sc iii}], and H$\rm \alpha$ emitters respectively.

Given the depth of our combined optical image, we can calculate the  minimum  observer frame equivalent 
width (EW$\rm _{min}$)
 that would be required for an emission line object to be a \lya\ emitter candidate.
We calculated the observer frame  EW using the following relation \citep{rho01}:
\begin{equation}
\rm EW_{min}\approx  \left [  \frac{f_{nb}}{f_{bb}}-1  \right ] \; \Delta \lambda_{nb} 
			=\left [ \frac{5 \sigma_{nb}}{2\sigma_{bb}}-1\right ]\; \Delta \lambda_{nb}
\end{equation}
where $\rm f_{nb} \; and \; f_{bb}$ are the fluxes in UNB and combined optical image 
respectively, $\rm \Delta \lambda_{nb}$ is the 
UNB filter width, and $ \rm \sigma_{nb} \; and \;  \sigma_{bb}$ are the uncertainties in flux measurements in UNB
and combined optical image respectively.  (The implicit approximation that the continuum contributes 
negligibly to 
the narrowband flux, is well justified for our 9\AA\ bandpass.)
With  $ \rm 5 \sigma_{nb} =7.8\times10^{-29} erg \;cm^{-2} s^{-1} Hz^{-1} \; and \;  2 \sigma_{bb}=
1.5 \times10^{-30} erg\; cm^{-2} s^{-1} Hz^{-1}$,
 we found that the foreground emission line sources would require $\rm EW_{min}\gtrsim 
460 \AA$ to contaminate our \lya candidate sample.\\

\textbf{ Foreground [O\,{\sc ii}]and [O\,{\sc iii}] emitters:}
Unfortunately, the equivalent width distribution of [O\,{\sc ii}] emitters has not been directly
measured at $z$=1.85. 
However,  several  authors \citep{tep03, kak07, str09} have studied [O\,{\sc ii}] emitters at $z<$1.5.
Here we use [O\,{\sc ii}] EW distribution,  obtained by  \citet{str09} at $\langle z\rangle  \approx$1
 in GOODS-south field, with the 
assumption that
 there is no significant evolution of the [O\,{\sc ii}] LF from $z$=1 to $z$=1.85.
In our \lya\ candidate selection, emission line sources with $I_{AB}$ fainter than 25.9 magnitude, and with 
$\rm EW_{obs}>460 \AA$ can contaminate our sample.
We determined which sources from  \citet{str09} would have passed these criteria
if redshifted to $z = 1.85$, and scaled the result by the ratio of volumes between the
two surveys.  We find that less than one (0.1)  
[OII]  emitter is expected to contaminate our \lya\ candidate sample.
 To be conservative, even if we relax the above magnitude cut by 0.5 mag to account for any color
correction, and  lower the 
$\rm EW_{obs}>200  \AA$, we find that less than 0.3 [O\,{\sc ii}] emitters should be expected to contaminate
our sample.

We apply a similar methodology to estimate the contamination from 
foreground [O\,{\sc iii}] emitters 
 \citep{kak07, hu09, str09, str10} 
at $\langle z \rangle \approx 1.1$ in our NEWFIRM
data using the [O\,{\sc iii}] emission line sources at  $\langle z \rangle=$0.5 in \citet{str09}.
%
 % An [OIII] emitter in \citet{str09} sample needs to have a line flux $ > 2.8\times10^{-17} \rm erg\; cm^{-2} s^{-1}$,
 % observed $\rm EW > 227 \AA$, and $I_{AB}$ fainter than 24.2 mag.
%  Applying these criteria and considering the volume difference between the two surveys,
 We found that less than  two (1.7) [O\,{\sc iii}] emitters  can be misidentified as \lya\ emitters
 in our survey.
 In addition to the above estimate,  we used a recent sample  of 
emission line galaxies  obtained from HST WFC3 early release science data \citep{str10}.
This sample of [O\,{\sc iii}]  emitters is closer in redshift, with median $z=1.1$,  to our foreground 
 [O\,{\sc iii}]  interloper redshift of $z=1.12$, thus minimizing the error in our  [O\,{\sc iii}] estimate
 due to possible evolution in the LF of  [O\,{\sc iii}] emitters.
 Using this recent sample, we found that about one  [O\,{\sc iii}] emitter is expected to contaminate our
 \lya\ candidate sample. \\

\textbf{ Foreground H${\alpha}$ emitters:}
As mentioned earlier, H${\alpha}$ emitters at $z$=0.62 can contaminate our \lya candidate sample.
Several authors \citep{tre02, str09} have studied H${\alpha}$ emitters at similar redshift.
\citet{tre02} (see their Figure 6) have plotted the \ha luminosity vs the continuum B-band magnitude of \ha emitters.
To pass our selection criteria,  an \ha emitter would require a luminosity  greater than $1\times10^{40} \rm erg 
\; s^{-1}$, and flux density $f_{B_{w}}< 7.5\times10^{-20} \rm erg\; cm^{-2} s^{-1} Hz^{-1}$  which corresponds to $\rm  M_{AB}=-15.97$ mag.
Any source brighter than  $\rm  M_{AB}$=-15.97 mag would be detected in the $B_{w}$ image, and hence rejected 
from \lya  candidate list.
From figure 6 \citep{tre02},  we expect to find no sources  that  can pass this selection criteria.
In addition, we used H${\alpha}$ emitters at $\langle z \rangle$=0.27  \citep{str09}, and found that less than one (0.4) H${\alpha}$ 
emitters are expected  to  contaminate our \lya\ candidate sample.

\subsection{Other Contaminants}
\textbf{Transient objects} : 
We rule out the possibility of contamination of our \lya\ candidates by transient objects such as 
supernovae, because  these objects would appear in both UNB and J band stacks.
Both UNB and J  data were obtained on each clear night of the run.\\

\textbf{L and T Dwarfs} :
Following  \citet{hib09} we  determined the expected number of L/T dwarfs in 
our survey. From the spectral type vs. absolute magnitude relations given by figure 9 in \cite{Tinney}, 
we infer that we could detect L dwarfs at a distance of 400 to 1300 pc
and T dwarfs at a distance of 150 to 600 pc, from
 the coolest to the warmest spectral types. 

Our field is located at a high galactic latitude, so that we would be able to detect L/T dwarfs 
well beyond the Galactic disk scale height.   
However, only a Galactic disk scale height of 350 pc is applicable to the 
population of L/T dwarfs \citep{Ryan}. We derive then a sampled volume of
$\sim$ 750 pc$^3$. Considering a volume density of L/T dwarfs of a few 
10$^{-3}$ pc$^{-3}$, we expect no more than one L/T dwarf in our field.

While we expect about one L/T dwarf in our survey, we further investigate if any of the observed
 L/T dwarf pass our selection criteria.
 To do this we selected about 160 observed spectra of L/T dwarf  
 \footnote{http://staff.gemini.edu/$\sim$sleggett/LTdata.html} \citep{gol04, kna04, chi06}, 
 and calculated the flux transmitted through the UNB and J-band filter.
 We found that  none of the  L/T dwarf has sufficient narrowband excess to pass our selection criteria.
 Therefore it is unlikely that our \lya\ candidate sample is contaminated by  L/T dwarf.\\

\textbf{Noise Spikes}:
Noise  in the detector can cause random flux increase in the UNB  filter.
To avoid contamination from such noise spikes, we constructed light curves of each candidate
using individual night stacks i.e. we selected candidates only if their flux was constant 
over all nights.
This method of candidate selection based on the constant flux in the individual night stacks also eliminates the 
possible contamination from persistence.\\

\textbf{Contribution from false detection:}
Finally, we performed a false detection test to estimate the number of false detection that
can  pass our \lya\ selection criteria.
To do this we multiplied the UNB stack by -1 and repeated the exact same procedure 
as the real \lya\ candidate selection (see section 3).
We did not get any false detection passing our selection criteria.

\section{Monte-Carlo Simulations}
Based on the above estimates   less than two [O\,{\sc iii}] emitters 
are expected to be misidentified as \lya\ emitters in our survey.
To estimate the number of sources that should be detected in our survey for a given \lya\ LF, 
we performed detailed Monte-Carlo simulations.
This is needed, since the width of the filter is comparable to or slightly smaller than the expected 
line width in these galaxies, so many of the sources will not be detected at their real line fluxes. 
In these simulations, we used the $z=6.6$  \lya\ LF derived by \citet{kas06}.

First, we generated  one million random galaxies distributed  according to the observed \lya\ 
LF at  $z$=6.6 \citep{kas06}. 
Each of these galaxies was assigned  a \lya\ luminosity in the range $\rm 1\times 10^{42} < Ly\alpha < 
1.5 \times 10^{43} erg \; s^{-1}$.
Here we assumed that the \lya\ LF does not evolve from $z$=6.6 to $z$=7.7.
Each galaxy was then assigned a random redshift $z_{L} < z < z_{H}$ where $z_{L}$ and $z_{H}$
correspond to the minimum and maximum 
wavelengths where the transmission of the UNB filter drops to zero.

Next, to each galaxy we assigned a flux $\rm F=L_{Ly\alpha}/4\pi d_{L}^{2}$ where $\rm d_{L}$ is the luminosity distance.  We distribute this flux in wavelength using an asymmetric \lya\  line profile
drawn from the $z=5.7$ spectra of  \citet{rho03}.
The flux transmitted through the UNB filter was then determined as $f_{trans} = \int f_\lambda T_\lambda 
d\lambda$ (where $T_\lambda$ is the filter transmission and $f_\lambda$ the flux density of the
emission line).  This accounts for the loss of the \lya\ flux that results from a filter whose width is
comparable to the line width (and not much greater as would be the case for a 1\% filter).
We then created a histogram of magnitudes after converting the convolved flux to magnitudes calculated using the following relation:
\begin{equation}
% \rm mag_{vega}=-2.5 \; Log_{10} \left( \it \frac{f_{trans} }{f_{0}} \right ),
\rm mag_{AB}=-2.5 \; Log_{10} \left( \it \frac{f_{trans} }{f_{0}} \right ),
\end{equation}
%where $f_{conv}$ is the convolved flux ($\rm erg \;s^{-1} cm^{-2}$), and 
and
\begin{equation}
%% f_{0}=\frac{ \rm 1.6\;kJy \times c }{(1.06\mu)^{2}} \times  \int T_{\lambda} \; d\lambda \; \; \; \rm erg \;s^{-1} cm^{-2},
f_{0}=\frac{ \rm 3.6\;kJy \times c }{(1.06\mu)^{2}} \times  \int T_{\lambda} \; d\lambda \; \; \; \rm erg \;s^{-1} cm^{-2},
\end{equation}
 with $c$ the speed of light.

Lastly, to include the instrumental effects, we multiplied the number of galaxies in each magnitude bin 
by the corresponding recovery fraction obtained from our artificial source simulations in 
our UNB image(see section 2.4).
We then converted each magnitude bin to a \lya\ luminosity bin, and counted the number of 
detected galaxies in each luminosity bin.

We repeated this simulation ten times, and taking an average, we found
that about one \lya\ emitter should be expected in our survey. 
It should be noted that we assumed a non-evolving  \lya\ LF from $z$=6.6 to $z$=7.7, 
and  that every \lya\ emitter has the same asymmetric \lya\ line profile.
While we expect about one \lya\ emitter in our survey there are large uncertainties
mainly due to the Poisson noise, and field to field variation or cosmic variance.
\citet{til09} have estimated field to field variation of \lya\ emitters to be $\gtrsim 30\%$ for a volume and 
flux limited \lya\ survey with a survey volume $\rm \sim 2 \times 10^{5} \; Mpc^{3}$.
We expect a larger field to field variation for smaller survey volumes.
We also estimated the cosmic variance expected in our survey using the cosmic
variance calculator \citep{tre08}.
For our survey  we should expect a cosmic variance of about $58\%$ assuming an intrinsic number of \lya\ sources
at $z=7.7$  in agreement with a non-evolving \lya\ LF from $z=6.6$ \citep{kas06} to $z=7.7$.
On the other hand our candidate counts are quite consistent with the
luminosity function at z=5.7 \citep{ouc09}.

\begin{figure}[t!]
\epsscale{1.2}
\plotone{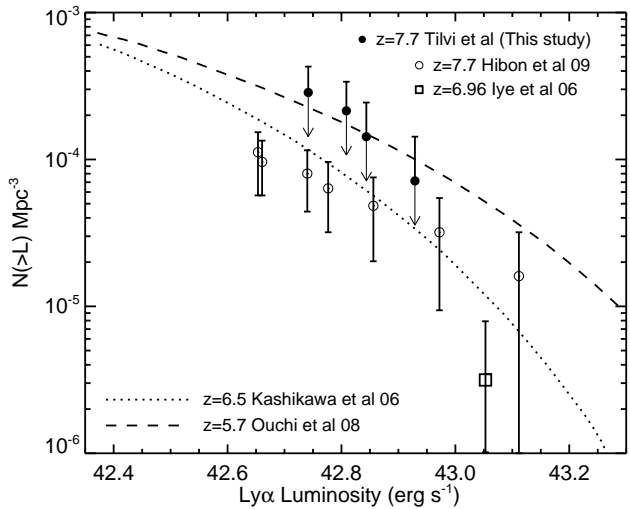}
\caption{Cumulative \lya\  luminosity function of our $z$=7.7 candidates (filled circles).
The filled points show the LF that will result if all four \lya\ galaxy candidates are confirmed.
The upper error bars are Poisson errors based on our sample size, while the down-arrows
below each data point indicate the possibility of a lower LF if some candidates are 
extreme emission line galaxies at lower redshift.
The open circles represent the LF from \citet{hib09} while the dashed line and dotted line show \lya\ LFs at $z$=5.7 \citep{ouc08} 
and $z$=6.5 \citep{kas06} respectively.
The open square is the LF at $z$=6.96 \citep{iye06}.
}
\end{figure}

\section{\lya\ luminosity function at $z$=7.7}
Using a large sample of \lya candidates, \citet{ouc08} found no significant evolution of \lya\ LF between
$z$=3.1 and $ z$=5.7. The evolution of  the \lya\ LF between $z=$5.7 
and $z=6.5$ is not conclusive.
For example, \citet{mr04} found no significant evolution of \lya\ LF between $z$=5.7 
and $z=6.5$, while \citet{kas06} suggest  an evolution of bright end of the LF in this 
redshift range.
On the theoretical front, several models  \citep{tho05,fer05, del06, dij07,kob07,mcq07, day08, nag08,sam09,til09} have been   developed 
to predict redshift evolution of  the \lya\ LF.
While several models  \citep[e.g.][]{sam09, til09} predict no significant evolution of \lya\ LF
 at  $z\lesssim 7$, the predictions differ greatly among different models.
These differences among the models can be attributed to differing input 
assumptions,  which in turn stem from our
imperfect understanding of the physical nature of \lya\ galaxies, and from 
the small samples currently available at high redshift.

\begin{center}
\begin{table*}[ht]
\caption{\lya\ searches at $z>$7.} 
{\small
\hfill{}
\begin{tabular}{l*{5}{c}r}
\tableline
$z$            & Survey   				& Detection limits 	& No. of LAE 		& Ref.\\
		& volume	(Mpc$^{3}$)		&	erg s$^{-1}$	& candidates		&	 \\
\hline
7.7 	& $1.4 \times 10^{4}$ 	& $6 \times 10^{-18} $ 		& 4		& This study  \\
7.7	& $6.3 \times 10^{4}$ 	& $8.3\times 10^{-18}$ 		& 7		& Hibon et al 2009   \\
8-10   &$35$				&$2\times 10^{-17}$			& 6		& Stark et al 2007 \\
8.8	& 3 arcmin$^{2}$		& $\sim 10^{-18} $			&0		& Parkes et al 1994\\
8.8	&$991$ 				& $2\times 10^{-17} $ 		& 0		& Willis et al 2005 \\
8.8	&$6.3 \times 10^{4}$ 	& $1.3\times 10^{-17}$	 	& 0 		& Cuby et al 2007 \\
8.96 & $1.12\times 10^6$  & $6\times 10^{-17} $ & 0 & Sobral et al 2009 \\
$\sim 9$  &$ \sim 450$ 		& $3.7\times 10^{-18}$	 	& 0 		& Willis et al 2008  \\
\hline
\end{tabular}}
\hfill{}
\label{tb:tablename}
\end{table*}
\end{center}

At $z>$ 6.5, there are only  a few searches for \lya\ emitters. 
\citet{iye06} found one spectroscopically confirmed LAE at $z$=6.96, and currently there are
no spectroscopically confirmed LAEs  at $z>$ 7.
However, there are few photometric searches \citep{par94, wil05, cub07, hib09} 
for \lya\ galaxies, and constraints
on \lya\ LF  at $z>$ 7.
Table 2 shows  details of different \lya\ searches at $z>$7. 

After careful selection of \lya\ candidates  and eliminating possible sources of 
contamination, we have found four \lya\ emitter candidates in a survey area of $28 \times 28$ arcmin$^{2}$,
with a limiting flux of $\rm 6 \times 10^{-18} erg \;s^{-1} \; cm^{-2}$ .
The fluxes of these four candidates are 1.1, 0.91, 0.84 and 0.72 in units of 
$\rm 10^{-17} \; erg \;s^{-1} \; cm^{-2}$.
Fig.\ 3 shows the resulting cumulative \lya\ luminosity function.
Solid filled circles show the \lya\ LF derived from our candidates, while open circles represent
\lya LF from \citet{hib09}. Arrows indicate that this is the upper limit on the \lya\ LF, and upper error
bars are the Poisson errors.
The dotted and dashed lines show \lya\ LFs from \citet{ouc08} and \citet{kas06} respectively. 
The open square is the \lya\ LF at $z$=6.96 \citep{iye06}.

If all of our \lya candidates are $z$=7.7 galaxies, the LF derived from our sample shows moderate
evolution compared to LF at $z$=6.5 \citep{kas06}. 
On the other hand, conservatively  if only one of the candidates
is a $z=7.7$ galaxy, then the \lya\ LF does not show any evolution compared to the $z=6.6$ \lya\ LF.
\citet{hib09}  conclude  that the observed \lya\ LF at $z=7.7$ does not evolve significantly 
compared to \lya\ LF at $z$=6.5 \citep{kas06}, if they consider that all of their candidates
are real. 
Finally, while our  \lya\ LF lies above the LF obtained by \citet{hib09}, 
the counts are consistent  with the  number of star-forming galaxies in the HUDF
with inferred \lya\ line fluxes $\rm > 6\times 10^{-18} erg \;s^{-1} \; cm^{-2}$
 \citep{fin09b}, and also consistent with the \lya\ LF at $z$=5.7 \citep{ouc08}.

As described in Section 5, all surveys for \lya\ emitters at $z > 6$ suffer from cosmic variance. We do
expect to see field-to-field variation in number counts even at the
same redshift. Therefore it is important to get statistics from more
than one field for each redshift. The field-to-field variation is
expected to be stronger for brighter sources. Therefore the higher
redshift surveys, which are more sensitivity limited, are hit the
hardest.

%\begin{tabular}{l*{4}{c}r}
%\hline
%$z$              & Survey   		& Detection limits & Ref.\\
%			& volume Mpc$^{3}$& erg s$^{-1}$ cm$^{2}$ & \\
%\hline
%7.7			& $6.3 \times 10^{4}$ & $8.3\times 10^{-18}$ & Hibon et al 2009  \\
%8.8			& 
%\end{tabular}

%\begin{center}
%\begin{table*}[ht]
%{\small
%\hfill{}
%\begin{tabular}{l*{2}{c}r}
%\hline
%$\#$            & Flux   			\\
%		& $\rm {erg  \; s^{-1} cm^{2}}$		\\
%\hline
%1	&  $1.1\times 10^{-17}  $ \\
%2 	&  $9.1\times 10^{-18}$  \\
%3	& $8.4 \times 10^{-18}$	\\
%4	& $7.2 \times 10^{-18}$	 \\
%\hline
%\end{tabular}}
%\hfill{}
%\label{tb:tablename}
%\end{table*}
%\end{center}

\section{\lya\ Equivalent Width}
Several studies have found numerous \lya emitters having large rest-frame  equivalent widths,
$\rm EW_{rest} > 240 \AA$
\citep{mr02, shi06, daw07, gro07,ouc08}.   These exceed theoretical predictions for
normal star forming galaxies.

Since the J-band filter does not include the \lya\ line, we have used the following relation to calculate the rest-frame 
\lya\ EWs for our four \lya\ candidates:
\begin{equation}
% \rm  EW_{rest}=\frac{f_{\lambda, NB}}{f_{\lambda, BB}}\times   \frac{1}{(1+z)} ~~.
\rm  EW_{rest}=\frac{f_{NB}}{f_{\lambda, BB}}\times   \frac{1}{(1+z)} ~~.
 \end{equation}
Here $\rm f_{ {NB}}$ and $\rm f_{ \lambda,{BB}}$ are the UNB line flux ($\rm erg \; s^{-1} cm^{2}$), and J-band flux (erg s$^{-1}$ cm$^{2}$ \AA$^{-1}$) respectively.
Since none of the four candidates are detected in J-band, we used J-band limiting magnitude to calculate
a lower limit on the \lya\ EWs.
We note that the  \lya  EW will depend on the exact redshift, shape, and precise position of the \lya line in the UNB filter.
However,  for simplicity and because we only put lower limits on EWs, we assume that the UNB filter encloses all the \lya line flux in calculating EWs.

% For our  \lya\ candidates having  fluxes 
% 4.1$ \times 10^{-29}$, $ 4.8\times10^{-29},$   $5.2\times  10^{-29}$, and   $ 6.3\times10^{-29}$
% in units of  erg s$^{-1}$ cm$^{2}$ Hz$^{-1}$, and with limiting magnitude of J-band $\rm J_{NF} =23.5$mag, 
% we obtained \lya\  $\rm EW_{rest} \gtrsim 4 \AA$ for our \lya candidates.

For our \lya\ candidates, with line flux estimates from $7$ to $11 \times 10^{-18} \ergcm2s$,
and our broad band limit $J_{NF} \ge 23.5$ mag, we find  
\lya\  $\rm EW_{rest} \gtrsim 3 \AA$.

This EW limit is considerably smaller than the $\rm EW_{rest} > 9 \AA$  obtained by 
\citet{hib09} for their  \lya\  candidates at $z$=7.7.
This difference  arises  due to the smaller bandwidth of our UNB
filter, and our somewhat shallower J band imaging.  Deep J-band observations will help 
in getting either measurements or stricter lower limits on the line EWs, 
but will also be observationally challenging.

\section{Summary and Conclusions}
We have performed a deep, wide field search for $z=$7.7 \lya\ emitters
on the NEWFIRM camera at the KPNO 4m Mayall telescope.  We used 
an ultra-narrowband filter with width 9$\rm \AA$ and central wavelength of 1.063$\mu m$,
yielding high sensitivity to narrow emission lines.

After careful selection of candidates by eliminating possible sources of contamination, we
detected four  candidate \lya\ emitters  with line flux  $\rm > 6 \times 10^{-18} erg \;s^{-1} \; cm^{-2}$ in a comoving volume of $\rm 1.4\times 10^{4} \; Mpc^{3}$.
%  The resulting \lya LF indicates that  there is only a moderate evolution  of \lya LF from $z$=6.5 to $z$=8. 
While we have carefully selected these four \lya\ candidates, we note that the 
number of \lya\  candidates is more than 
the expected number obtained by using the z=6.6 luminosity function of
Kashikawa et al. 2006, though quite consistent with the z=5.7 luminosity
function of \citet{ouc08}.  Hence, our results would allow for a modest {\it increase} 
in the \lya\ LF from $z=6.5$ to $z\approx 8$.  Spectroscopic confirmation of more than
two candidates would show that such an increase is in fact required.
However,  more surveys are needed to account for the uncertainty due to cosmic variance.

In order to use the \lya\ luminosity functions as a test of
reionization, we need to be able to detect variations in $ L^{\star}$, 
the characteristic luminosity, of factors of
three or four.  This will require larger samples, spectroscopic
confirmations, and a measure of field-to-field variation.

It is therefore premature to draw any conclusions about reionization
from the current sample. It is, however, encouraging that we are able
to reach the sensitivity and volume required to detect multiple
candidates robustly.

\acknowledgments
We thank the anonymous referee for insightful comments and suggestions, 
and  thank the staff of the KPNO for their support. ÊWe also thank Buell Jannuzi, Ilian T. Iliev, Bahram Mobasher, 
Hyron Spinrad, Arjun Dey, and Norbert Pirzkal for helpful discussions in the course of this work. We gratefully 
acknowledge financial support from the National Science Foundation through NSF grants AST-0808165 and 
AST-0606932.


\begin{thebibliography}{}

\bibitem[Ajiki et al.(2004)]{aji04} Ajiki, M., et al.\ 2004, 
\pasj, 56, 597 



\bibitem[Autry et al.(2003)]{aut03} Autry, R.~G., et al.\ 
2003, \procspie, 4841, 525 

\bibitem[Bertin 
\& Arnouts(1996)]{ber96} Bertin, E., \& Arnouts, S.\ 1996, \aaps, 117, 393 

\bibitem[Bertin et al.(2002)]{ber02} Bertin, E., Mellier, Y., 
Radovich, M., Missonnier, G., Didelon, P., 
\& Morin, B.\ 2002, Astronomical Data Analysis Software and Systems XI, 281, 228 

\bibitem[Chiu et al.(2006)]{chi06} Chiu, K., Fan, X., 
Leggett, S.~K., Golimowski, D.~A., Zheng, W., Geballe, T.~R., Schneider, 
D.~P., \& Brinkmann, J.\ 2006, \aj, 131, 2722 


\bibitem[Cowie 
\& Hu(1998)]{cow98} Cowie, L.~L., \& Hu, E.~M.\ 1998, \aj, 115, 1319 


\bibitem[Cuby et 
al.(2007)]{cub07} Cuby, J.-G., Hibon, P., Lidman, C., Le F{\`e}vre, O., Gilmozzi, R., Moorwood, A., 
\& van der Werf, P.\ 2007, \aap, 461, 911 

\bibitem[Dawson et al.(2004)]{daw04} Dawson, S., et al.\ 
2004, \apj, 617, 707 


\bibitem[Dawson et al.(2007)]{daw07} Dawson, S., Rhoads, 
J.~E., Malhotra, S., Stern, D., Wang, J., Dey, A., Spinrad, H., 
\& Jannuzi, B.~T.\ 2007, \apj, 671, 1227 

\bibitem[Dayal et al.(2008)]{day08} Dayal, P., Ferrara, A., 
\& Gallerani, S.\ 2008, \mnras, 389, 1683 



\bibitem[Dickinson \&  Valdes (2009)]
{dic09}  Dickinson, M. \& Valdes,  F.  A Guide to NEWFIRM Data Reduction with IRAF, NOAO SDM 
 PL017, 2009


\bibitem[Dijkstra et al.(2007)]{dij07} Dijkstra, M., Wyithe, 
J.~S.~B., \& Haiman, Z.\ 2007, \mnras, 379, 253 


%\bibitem[Ellis 
%\& Bland-Hawthorn(2008)]{ell08} Ellis, S.~C., \& Bland-Hawthorn, J.\ 2008, \mnras, 386, 47 


\bibitem[Furlanetto et al.(2005)]{fer05} Furlanetto, S.~R., 
Schaye, J., Springel, V., \& Hernquist, L.\ 2005, \apj, 622, 7 

\bibitem[Furlanetto et al.(2006)]{fur06} Furlanetto, S.~R., 
Zaldarriaga, M., \& Hernquist, L.\ 2006, \mnras, 365, 1012 



\bibitem[Finkelstein et al.(2007)]{fin07} Finkelstein, S.~L.,  Rhoads, J.~E., Malhotra, S., Pirzkal, N., 
\& Wang, J.\ 2007, \apj, 660, 1023

\bibitem[Finkelstein et al.(2009)]{fin09} Finkelstein, S.~L., 
Rhoads, J.~E., Malhotra, S., \& Grogin, N.\ 2009, \apj, 691, 465 


\bibitem[Finkelstein et al.(2009b)]{fin09b} Finkelstein, S.~L., 
Papovich, C., Giavalisco, M., Reddy, N.~A., Ferguson, H.~C., Koekemoer, 
A.~M., \& Dickinson, M.\ 2009, arXiv:0912.1338 


\bibitem[Fynbo et 
al.(2001)]{fyn01} Fynbo, J.~U., M{\"o}ller, P., \& Thomsen, B.\ 2001, \aap, 374, 443 



\bibitem[Gawiser et al.(2006)]{gaw06} Gawiser, E., et al.\ 2006, \apjl, 642, L13

\bibitem[Golimowski et al.(2004)]{gol04} Golimowski, D.~A., 
et al.\ 2004, \aj, 127, 3516 


\bibitem[Gronwall et al.(2007)]{gro07} Gronwall, C., et al.\ 
2007, \apj, 667, 79

\bibitem[Hibon et al.(2009)]{hib09} Hibon, P., et al.\ 2009, 
arXiv:0907.3354

\bibitem[Horton et al.(2004)]{hor04} Horton, A., Parry, I., 
Bland-Hawthorn, J., Cianci, S., King, D., McMahon, R., 
\& Medlen, S.\ 2004, \procspie, 5492, 1022 

\bibitem[Hu et al.(1999)]{hu99} Hu, E.~M., McMahon, R.~G.,
\& Cowie, L.~L.\ 1999, \apjl, 522, L9

\bibitem[Hu et al.(2002)]{hu02} Hu, E.~M., Cowie, L.~L., 
McMahon, R.~G., Capak, P., Iwamuro, F., Kneib, J.-P., Maihara, T., 
\& Motohara, K.\ 2002, \apjl, 568, L75 

\bibitem[Hu et al.(2004)]{hu04} Hu, E.~M., Cowie, L.~L., 
Capak, P., McMahon, R.~G., Hayashino, T., 
\& Komiyama, Y.\ 2004, \aj, 127, 563 

\bibitem[Hu et al.(2009)]{hu09} Hu, E.~M., Cowie, L.~L., 
Kakazu, Y., \& Barger, A.~J.\ 2009, \apj, 698, 2014 



\bibitem[Iye et al.(2006)]{iye06} Iye, M., et al.\ 2006,  \nat, 443, 186

\bibitem[Jannuzi 
\& Dey(1999)]{jan99} Jannuzi, B.~T., \& Dey, A.\ 1999, Photometric Redshifts and the Detection of High 
Redshift Galaxies, 191, 111 

\bibitem[Kakazu et al.(2007)]{kak07} Kakazu, Y., Cowie, 
L.~L., \& Hu, E.~M.\ 2007, \apj, 668, 853 


\bibitem[Kashikawa et al.(2006)]{kas06} Kashikawa, N., et  al.\ 2006, \apj, 648, 7 

\bibitem[Knapp et al.(2004)]{kna04} Knapp, G.~R., et al.\ 
2004, \aj, 127, 3553

\bibitem[Kobayashi et al.(2007)]{kob07} Kobayashi, M.~A.~R., 
Totani, T., \& Nagashima, M.\ 2007, \apj, 670, 919

\bibitem[Kodaira et al.(2003)]{kod03} Kodaira, K., et al.\ 
2003, \pasj, 55, L17 



\bibitem[Komatsu et al.(2010)]{kom10} Komatsu, E., et al.\ 
2010, arXiv:1001.4538 

\bibitem[Kudritzki et al.(2000)]{kud00} Kudritzki, R.-P., et 
al.\ 2000, \apj, 536, 19 



\bibitem[Le Delliou et al.(2006)]{del06} Le Delliou, M., 
Lacey, C.~G., Baugh, C.~M., \& Morris, S.~L.\ 2006, \mnras, 365, 712 


\bibitem[Maihara et al.(1993)]{mai93} Maihara, T., Iwamuro, 
F., Yamashita, T., Hall, D.~N.~B., Cowie, L.~L., Tokunaga, A.~T., 
\& Pickles, A.\ 1993, \pasp, 105, 940

\bibitem[Malhotra 
\& Rhoads(2002)]{mr02} Malhotra, S., \& Rhoads, J.~E.\ 2002, \apjl, 565, L71 


\bibitem[Malhotra 
\& Rhoads(2004)]{mr04} Malhotra, S., \& Rhoads, J.~E.\ 2004, \apjl, 617, L5 

\bibitem[Malhotra 
\& Rhoads(2006)]{mr06} Malhotra, S., \& Rhoads, J.~E.\ 2006, \apjl, 647, L95 

\bibitem[McQuinn et al.(2007)]{mcq07} McQuinn, M., Hernquist, 
L., Zaldarriaga, M., \& Dutta, S.\ 2007, \mnras, 381, 75 



\bibitem[Nagamine et al.(2008)]{nag08} Nagamine, K., Ouchi, 
M., Springel, V., \& Hernquist, L.\ 2008, arXiv:0802.0228 


\bibitem[Nilsson et 
al.(2007)]{nil07} Nilsson, K.~K., et al.\ 2007, \aap, 471, 71 



\bibitem[Ota et al.(2008)]{ota08} Ota, K., et al.\ 2008, 
\apj, 677, 12 


\bibitem[Ouchi et al.(2001)]{ouc01} Ouchi, M., et al.\ 2001, 
\apjl, 558, L83 

\bibitem[Ouchi et al.(2003)]{ouc03} Ouchi, M., et al.\ 2003, 
\apj, 582, 60 

\bibitem[Ouchi et al.(2008)]{ouc08} Ouchi, M., et al.\ 2008, 
\apjs, 176, 301

\bibitem[Ouchi et al.(2009)]{ouc09} Ouchi, M., et al.\ 2009, 
\apj, 696, 1164 

\bibitem[Papovich et al.(2001)]{pap01} Papovich, C., 
Dickinson, M., \& Ferguson, H.~C.\ 2001, \apj, 559, 620 



\bibitem[Parkes et al.(1994)]{par94} Parkes, I.~M., Collins, 
C.~A., \& Joseph, R.~D.\ 1994, \mnras, 266, 983 

\bibitem[Pentericci et 
al.(2000)]{pen00} Pentericci, L., et al.\ 2000, \aap, 361, L25 


\bibitem[Pentericci et al.(2009)]{pen09} Pentericci, L., Grazian, A., Fontana, A., Castellano, M., Giallongo, E., Salimbeni, S., \& Santini, P.\ 2009, \aap, 494, 553


\bibitem[Pirzkal et al.(2007)]{pir07} Pirzkal, N., Malhotra,
S., Rhoads, J.~E., \& Xu, C.\ 2007, \apj, 667, 49



\bibitem[Ryan et al.(2005)]{Ryan} Ryan, R.~E., Jr., Hathi, 
N.~P., Cohen, S.~H., \& Windhorst, R.~A.\ 2005, \apjl, 631, L159 

\bibitem[Rhoads(2000a)]{rho00a} Rhoads, J.~E.\ 2000, \pasp,  112, 703 

\bibitem[Rhoads et al.(2000b)]{rho00b} Rhoads, J.~E., Malhotra, 
S., Dey, A., Stern, D., Spinrad, H., 
\& Jannuzi, B.~T.\ 2000, \apjl, 545, L85

\bibitem[Rhoads 
\& Malhotra(2001)]{rho01} Rhoads, J.~E., \& Malhotra, S.\ 2001, \apjl, 563, L5 


\bibitem[Rhoads et al.(2003)]{rho03} Rhoads, J.~E., et al.\ 
2003, \aj, 125, 1006 

\bibitem[Rhoads et al.(2004)]{rho04} Rhoads, J.~E., et al.\ 
2004, \apj, 611, 59 


\bibitem[Rousselot et 
al.(2000)]{rou00} Rousselot, P., Lidman, C., Cuby, J.-G., Moreels, G., \& Monnet, G.\ 2000, \aap, 354, 1134 



\bibitem[Samui et al.(2009)]{sam09} Samui, S., Srianand, R., 
\& Subramanian, K.\ 2009, \mnras, 398, 2061 

\bibitem[Shapley et al.(2001)]{sha01} Shapley, A.~E., 
Steidel, C.~C., Adelberger, K.~L., Dickinson, M., Giavalisco, M., 
\& Pettini, M.\ 2001, \apj, 562, 95 

\bibitem[Shimasaku et al.(2006)]{shi06} Shimasaku, K., et 
al.\ 2006, \pasj, 58, 313

\bibitem[Sobral et al (2009)]{sob09} Sobral, D., et al 2009, \mnras, 398, L68
  % If you do a silly survey, you get no objects.

\bibitem[Stark et al.(2007)]{sta07} Stark, D.~P., Ellis, 
R.~S., Richard, J., Kneib, J.-P., Smith, G.~P., 
\& Santos, M.~R.\ 2007, \apj, 663, 10 

\bibitem[Stark et al.(2009)]{sta09} Stark, D.~P., Ellis, 
R.~S., Bunker, A., Bundy, K., Targett, T., Benson, A., 
\& Lacy, M.\ 2009, \apj, 697, 1493 


\bibitem[Steidel et al.(1996)]{ste96} Steidel, C.~C., 
Giavalisco, M., Pettini, M., Dickinson, M., 
\& Adelberger, K.~L.\ 1996, \apjl, 462, L17 

\bibitem[Stern et al.(2005)]{ste05} Stern, D., Yost, S.~A., 
Eckart, M.~E., Harrison, F.~A., Helfand, D.~J., Djorgovski, S.~G., 
Malhotra, S., \& Rhoads, J.~E.\ 2005, \apj, 619, 12 

\bibitem[Stiavelli et al.(2001)]{sti01} Stiavelli, M., 
Scarlata, C., Panagia, N., Treu, T., Bertin, G., 
\& Bertola, F.\ 2001, \apjl, 561, L37 



\bibitem[Straughn et al.(2009)]{str09} Straughn, A.~N., et 
al.\ 2009, \aj, 138, 1022

\bibitem[Straughn et al.(2010)]{str10} Straughn, A.~N., et 
al.\ 2010, arXiv:1005.3071 

\bibitem[Szalay et al.(1999)]{sza96} Szalay, A.~S., Connolly, 
A.~J., \& Szokoly, G.~P.\ 1999, \aj, 117, 68 


\bibitem[Taniguchi et al.(2005)]{tan05} Taniguchi, Y., et 
al.\ 2005, \pasj, 57, 165

\bibitem[Teplitz et al.(2003)]{tep03} Teplitz, H.~I.,  Collins, N.~R., Gardner, J.~P., Hill, R.~S., 
\& Rhodes, J.\ 2003, \apj, 589, 704

\bibitem[Thommes 
\& Meisenheimer(2005)]{tho05} Thommes, E., \& Meisenheimer, K.\ 2005, \aap, 430, 877 

\bibitem[Tilvi et al.(2009)]{til09} Tilvi, V., Malhotra, S., 
Rhoads, J.~E., Scannapieco, E., Thacker, R.~J., Iliev, I.~T., 
\& Mellema, G.\ 2009, \apj, 704, 724


\bibitem[Tinney et al.(2003)]{Tinney} Tinney, C.~G., 
Burgasser, A.~J., \& Kirkpatrick, J.~D.\ 2003, \aj, 126, 975 

\bibitem[Tokunaga et al.(2002)]{tok02} Tokunaga, A.~T., 
Simons, D.~A., \& Vacca, W.~D.\ 2002, \pasp, 114, 180 

\bibitem[Trenti 
\& Stiavelli(2008)]{tre08} Trenti, M., \& Stiavelli, M.\ 2008, \apj, 676, 767 


\bibitem[Tresse et al.(2002)]{tre02} Tresse, L., Maddox, 
S.~J., Le F{\`e}vre, O., \& Cuby, J.-G.\ 2002, \mnras, 337, 369 




\bibitem[Valdes(1998)]{val98} Valdes, F.~G.\ 1998, 
Astronomical Data Analysis Software and Systems VII, 145, 53 

\bibitem[Venemans et 
al.(2004)]{ven04} Venemans, B.~P., et al.\ 2004, \aap, 424, L17 



\bibitem[Wang et al.(2004)]{wan04} Wang, J.~X., et al.\ 2004,  \apjl, 608, L21 

\bibitem[Wang et al.(2007)]{wan07} Wang, J.~X., Zheng, Z.~Y., 
Malhotra, S., Finkelstein, S.~L., Rhoads, J.~E., Norman, C.~A., 
\& Heckman, T.~M.\ 2007, \apj, 669, 765 

\bibitem[Wang et al.(2009)]{wan09} Wang, J.-X., Malhotra, S., 
Rhoads, J.~E., Zhang, H.-T., \& Finkelstein, S.~L.\ 2009, \apj, 706, 762 



\bibitem[Willis 
\& Courbin(2005)]{wil05} Willis, J.~P., \& Courbin, F.\ 2005, \mnras, 357, 1348 

\bibitem[Willis et al.(2008)]{wil08} Willis, J.~P., Courbin, 
F., Kneib, J.-P., \& Minniti, D.\ 2008, \mnras, 384, 1039 


\end{thebibliography}
\end{document}